\documentstyle[preprint,aps]{revtex}
\begin{document}
\draft
\preprint{hep-th/9505108; IASSNS-HEP-95/34 ;   OUTP-95-20P}
\title{Wavefunctions for Non-Abelian Vortices}
\author{Kai-Ming Lee\footnote{kmlee@thphys.ox.ac.uk}}
\address{Department of Physics, University of Oxford,\\
 1 Keble Road, Oxford, OX1 3NP, U.K.}
\author{and}
\author{Hoi-Kwong Lo\footnote{hkl@sns.ias.edu}}
\address{
 School of Natural Sciences, Institute for Advanced Study, Olden Lane,\\
 Princeton, NJ 08540, U.S.A.}
\maketitle

\begin{abstract}
We construct exact wavefunctions of two
vortices on a plane, a single vortex on the cylinder and
a vortex on the torus. In each case, the physics
is shown to be equivalent
to a particle moving in a covering space, something simple
to solve in those examples.
We describe how our solutions fit
into the general theory of quantum mechanics
of $N$ particles on a two-dimensional space
and attribute our success to the fact that the fundamental
groups are Abelian in those simple cases that we are considering.
\end{abstract}


Vortices\cite{Vilenkin} are point-like topological defects on a two
dimensional space.
Consider the spontaneous symmetry breakdown of a simply-connected
gauge group
$K$ into a finite subgroup $G$. This
will give rise to stable topological vortices.\footnote{Vortices
arise whenever the low energy gauge group of a symmetry broken
theory has {\em disconnected} components. For simplicity,
we take the low energy gauge group to be finite throughout this
paper.}
Neglecting any finite-core effects, the low energy theory is
essentially a topological field theory with a finite gauge
group $G$ \cite{Witten,KraWil,Bais,Freed}.
A vortex carries a ``flux'' which can be labeled by an element of $G$.
If $G$ is non-Abelian, vortices can exhibit topological
interaction with one another: Adiabatically bringing a
vortex around another will change the fluxes of both
vortices even though they never come close to
each other \cite{Poe,Bucher,PresLo}.
Even in the case of a single vortex, non-trivial
topological interactions can still occur if the space
is {\em not} simply-connected \cite{PreKra,wormhole,Lee}.
A multiply-connected space
contains non-contractible loops. The flux of a vortex
may change when it traverses those loops.

Both the vortex-vortex interactions and the interactions between
a vortex and a non-contractible loop in space can be described
by flat connections in configuration spaces. As is well-known,
a flat connection can be trivialized at the expense of introducing
multivalued wavefunctions (i.e., wavefunctions
with non-trivial monodromy properties). This is the
strategy we are adopting in this paper.
We will find out those boundary conditions and construct
exact wavefunctions for
simple examples (two vortices on a plane and  a single vortex
on a cylinder or a torus). 
We also 
describe how our solutions fit into the general theory of
the quantum mechanics of $N$ particles on a two-dimensional
space. Finally, we remark that our success in solving those
simple cases is partly due to the fact that the fundamental
groups of their configuration spaces are Abelian.

Consider the vortices
that arise due to the symmetry breakdown of a gauge group into a finite
group $G$.
We assign an element of $G$ to any isolated vortex to label its flux
by the following method. Choose a fixed but {\em arbitrary} base point,
$x_0$, and a loop $C$ that encloses the vortex and begins and ends at
the point $x_0$. Associate the vortex with the {\em untraced}
Wilson loop operator:
\begin{equation}
a(C,x_0)= P \exp \left( i \int_{C,x_0} A dx \right) ~,\label{Wilson}
\end{equation}
where $P$ denotes the path ordering. Since the gauge field is massive
for a finite unbroken group, the
element $a(C,x_0)$ is invariant under deformations of the path $C$
that keep $x_0$ fixed and that avoid the vortex core. An object
that transforms as an irreducible representation $\nu$ of $G$
acquires an ``Aharonov-Bohm'' phase $D^{\nu}(a (C,x_0))$ when
covariantly transported around the vortex.
$a(C,x_0)$ has to be an element of $G$ because the Higgs condensate
must be invariant when so transported.

If there are two or more vortices, we must choose a standard loop
for each vortex as in FIG. 1. Then we assign
group elements $a_1, a_2 , \cdots, a_n$ to
the loops $\gamma_1, \gamma_2, \cdots, \gamma_n$ respectively.
This description is ambiguous because under a gauge
transformation by $g \in G$ at the base point $x_0$, the
elements $a_1, a_2 , \cdots, a_n$
transform according to
$a_i \to g a_i g^{-1}~.$
For a single vortex, the gauge transformations
act transitively on the conjugacy class of $G$
to which a vortex belongs. Thus, one
might be tempted to say that the flux of a vortex should
really be labeled by a conjugacy class rather than a group element.
But this is not correct because there is only one overall
gauge degree of freedom. If there are two vortices, labeled by group elements
$a$ and $b$ with respect to the same base point $x_0$, then the effect
of a gauge transformation at $x_0$ is
$g: a \to g a g^{-1} , b \to g b g^{-1}.$
Thus, if $a$ and $b$ are distinct representatives of the same
class, they remain distinct in any gauge.

We consider only the non-relativistic quantum mechanics
of non-interacting
non-Abelian vortices. {\em Locally}, the Hamiltonian is just
that of non-interacting particles
\begin{equation}
  H= \sum_{i=1}^{N} {p_i^2\over 2m_i}~.\label{ham}
\end{equation}
The complexity of the problem,
however, lies
in the non-trivial monodromy properties that the
wavefunctions must satisfy.

The case of a single vortex on an infinite plane is trivial because
there is no topological interaction and the vortex behaves
just like a free particle.
A more interesting example is two vortices on a plane.
The center of mass motion can be separated from the relative
motion.
Passing to the center of mass frame, the relative motion
is equivalent to that of a particle moving on the plane with
the origin deleted.\footnote{We are imposing the hard-core condition
that no two vortices
can coincide.}

The wavefunctions for the relative motion have a non-trivial
monodromy property. Standard paths $\gamma_1$ and $\gamma_2$ that
wind counterclockwise around the two vortices have
been chosen in FIG.~2a. Suppose now that vortex 1 winds around vortex 2
as in FIG.~2b. We may deform our paths during the winding so that no
vortex ever crosses any path; then each path is mapped to the same
group element after the winding as before the winding. But after the
winding, the final deformed path is not homotopically equivalent to
the initial path.

Suppose, for example, that initially $\gamma_1$ ($\gamma_2$
respectively) is mapped to $a_1$ ($a_2$ respectively).
To determine the final values, after the winding,
of the group elements associated with the paths $\gamma_1$ and
$\gamma_2$, notice that during the winding, the path
shown in FIG.~2c is ``dragged'' into $\gamma_1$. Hence, the group
element associated with this path before the winding will become the
element associated with $\gamma_1$ after the winding. Since this
path is homotopically equivalent to $(\gamma_1 \gamma_2) \gamma_1
(\gamma_1 \gamma_2)^{-1}$, where $\gamma_1 \gamma_2$ denotes the path
that is obtained by traversing $\gamma_2$ first and $\gamma_1$ second,
before the winding, the path is associated with
the element
\begin{equation}
a'_1 = (a_1 a_2) a_1 (a_1 a_2)^{-1} ~.\label{newfluxa}
\end{equation}
Similarly,
the path shown in FIG.~2d is dragged during the winding to $\gamma_2$.
This path is $(\gamma_1 \gamma_2) \gamma_2 (\gamma_1 \gamma_2)^{-1}$ and,
before the winding,
it is mapped to
\begin{equation}
a'_2 = (a_1 a_2 ) a_2 (a_1 a_2)^{-1} ~.\label{newfluxb}
\end{equation}
We conclude that,
when a vortex of flux $a$ 
winds counterclockwise
around a vortex of flux $b$ as shown in FIG. 2, the fluxes of
both vortices will be conjugated by $ab$ \cite{Poe,Bucher,PresLo}:
Denoting the
fluxes of the two vortices before the winding by $|a,b\rangle$,
on winding, 
\begin{equation}
  |a,b\rangle\to |(ab)a(ab)^{-1},(ab)b(ab)^{-1}\rangle~.\label{interchange}
\end{equation}

If the two vortices wind around each other $m$ times,
from (\ref{interchange}), their fluxes become
\begin{equation}
  |(ab)^m a (ab)^{-m}, (ab)^m b (ab)^{-m}\rangle~.\label{three}
\end{equation}
Since the unbroken group
$G$ is assumed to be
finite, the fluxes eventually return to their original values,
say after $n$ windings, i.e.,
$(ab)^n a (ab)^{-n}=a$ and $(ab)^n b (ab)^{-n}=b$. Let us fix the
$a$ vortex at the origin.
If the $b$ vortex goes around the origin $n$ times, the relative
wavefunction will not change at all. Notice that in polar
coordinates $(r,\theta)$, the usual requirement of
a periodicity of $2 \pi$ for $\theta$ not longer applies.
Owing to topological interactions, the required period
of the relative wavefunction is
$2 \pi n$ rather than $2 \pi$ \cite{PresLo}. (From the point of
view of a vortex, the physical space is an $n$-sheeted surface
\cite{BuGold} with
an $n$ that depends on $a$ and $b$.)

One may still attempt to restrict $\theta$ to the range between
$0$ and $ 2 \pi$.
If we denote
the state of the fluxes of the two vortices after $k$
windings by
\begin{equation}
|k \rangle = |(ab)^k a (ab)^{-k}, (ab)^k b (ab)^{-k}\rangle ,
\label{kwinding}
\end{equation}
at each angle $\theta$, the two vortices
can be in one of the $n$ flux eigenstates $|0 \rangle, |1 \rangle, \cdots,
|n-1 \rangle .$ Therefore, a wavefunction is represented by
a column vector with $n$ entries $\psi_0, \psi_1, \cdots, \psi_{n-1} $
satisfying the relations
\begin{equation}
\psi_k (r, \theta + 2 \pi) = \psi_{k+1} (r, \theta)~. \label{period}
\end{equation}

It is convenient to transform to the ``monodromy eigenstates.''
A basis vector
\begin{equation}
  \chi_l(r, \theta) ={1\over \sqrt n} \sum_{k=0}^{n-1} e^{-2\pi i kl /n}
  \psi_k(r, \theta)~.\label{five}
\end{equation}
{}From Eq.~(\ref{period}), $\chi_l$ has the property that
\begin{equation}
  \chi_l (r, \theta + 2 \pi) = e^{2\pi il /n} \chi_l ( r, \theta)~.\label{mono}
\end{equation}
As discussed in
Ref.~\cite{PresLo},
these monodromy eigenstates correspond to states of the two-vortex system
that have definite charge, in the sense that they are eigenstates of the gauge
transformation $ab \in G$, where $ab$ is the total flux.\footnote{The
wavefunctions of two charge-flux composites (i.e., dyons) can
similarly be obtained. The only difference is that, after $n$ windings,
the two-dyon state will return to itself only up to a phase, which
will change the eigenvalues of the monodromy operator.}

If two vortices are in the same conjugacy classes, the two vortices
should be regarded as indistinguishable\cite{PresLo}.
We should consider braiding
${\cal R}$
(counterclockwise exchange) between the two instead of winding.
It is possible that the braid operator has an orbit of odd order
acting on the two vortex state. In that case,
the wavefunction has a periodicity $ (2 n+1) \pi$.

We now turn to the case of a single vortex on the cylinder.
This case is similar to the previous one. Let us
choose the arbitrary base point to be at ``spatial infinity''.
Choose a standard path $\gamma$ around the vortex.
There is a
homotopy class of non-contractible loops on a cylinder.
Call it $\alpha$. Suppose the vortex winds along $\alpha$ as
shown in FIG.~3a. 
The standard path $\gamma$ will be dragged into a final deformed
path $\gamma'= \alpha^{-1} \gamma \alpha$ in FIG.~3b.
If initially $\alpha$ and $\gamma$ are assigned with fluxes
$a$ and $b$, let us denote this assignment by $|a,b \rangle$.
After the winding, the relation $\gamma'= \alpha^{-1} \gamma \alpha$
implies that the assignment changes to
$|a,aba^{-1} \rangle$.

Again since $G$ is finite, there is a minimal positive integer $n$ such that
$a^n b a^{-n}=b $. We can
proceed in the same manner as in the
case of two vortices on a plane and see that the quantum mechanics of a vortex
on a cylinder of radius $R$ is equivalent to that a free particle on
a cylinder with a larger radius $nR$.
A cylinder is diffeomorphic to an infinite
plane with the origin deleted. Therefore, a vortex on a
cylinder is topologically equivalent to two vortices on a plane.
This is the reason why the physics is essentially the same in the two cases.
One also notes that there
are two seperated infinite regions on the cylinder. Mapping
either of the two to the origin of the plane gives equivalent
results.

The last case that we will study is a vortex on the
torus. Once again, let us choose an arbitrary base point.
There are two homotopy classes of non-contractible loops on
a torus. We denote them by $\alpha$ and $\beta$. There are magnetic
fluxes associated with the two loops, $\alpha\mapsto a$ and
$\beta\mapsto b$. Let the flux of the vortex measured along
the path $\gamma$ be $c$.
Let us denote the state of the fluxes of the vortex and the two
non-contractible loops by
$|c;a,b\rangle$. Since the space is compact, the fluxes $a$, $b$ and
$c$ satisfy a relation \cite{Lee}. In the convention of
FIG~4a, the relation is
\begin{equation}
  c=b^{-1}a^{-1}ba~.\label{eleven}
\end{equation}
If the vortex goes
around the $\alpha$ loop, it is easy to see in FIG.~4b that the
paths have been smoothly deformed into
\begin{equation}
 \alpha' =\alpha~,~~~ 
 \beta' = \alpha^{-1} \beta \alpha~,~~~
 \gamma' = \alpha^{-1} \gamma \alpha  ~. \label{toruspaths}
\end{equation}
As the elements assigned to the deformed paths $\alpha'$, $\beta'$ and
$\gamma'$ will still be $a$, $b$ and $c$, we find that the elements
associated with the standard paths $\alpha$, $\beta$ and $\gamma$ will
be modified to
\begin{equation}
 a' = a~,~~~
 b' = aba^{-1}~,~~~
 c' = aca^{-1} ~. \label{torusflux}
\end{equation}
Notice that the effect of the winding is equivalent to a global
gauge transformation by the element $a$. Because a torus is compact,
if we regard our theory
on the two-dimensional torus
as fundamental, the Gauss law constraint for a compact surface demands that
the state of the whole torus be invariant under global gauge
transformations: A {\em closed} universe cannot carry any net gauge charges.
An example of a state that satisfies the Gauss law constraint is,
up to normalization,
\begin{equation}
\sum_{g\in G}|gcg^{-1}; gag^{-1}, gbg^{-1}\rangle~.\label{torusstate}
\end{equation}
The winding of the vortex around the $\alpha$ loop will, therefore, lead to
no observable changes. A similar argument applies to the winding around the
$\beta$ loop. In conclusion, the quantum mechanics of a vortex on a torus
is equivalent to that of a free particle on a torus of the same size.

Incidentally, our solution also resolves some complication concerning
the base point. It was noted \cite{Bucher} that, in
an $n$-vortex configuration, when a vortex winds around the
base point, the fluxes of all the vortices appear to be conjugated. However,
if
we regard the base point as arbitrary, winding around it should lead
to no observable changes. To
avoid this complication, 
it is convenient to place the base point at ``spatial infinity''.
However, for a compact surface like a torus, there is no
``spatial infinity'' to talk about.
One can no longer ignore the possibility of the winding of a vortex
around the base point on a torus. This will lead to
a gauge transformation of the whole
configuration. But there is an easy way out:
the Gauss law for a compact surface precisely
demands that all states
related by gauge transformations to be identified. Hence,
the base point is indeed arbitrary and winding around it leads
to no observable changes.

We will now describe how our results for those special cases
fit into the general theory of quantum mechanics of $N$ particles
on a two-dimensional space\cite{Bal}. In general discussions of
the quantum mechanics of $N$ particles, the following
framework is usually adopted: Suppose that the position of
each particle takes value in a manifold $M$.
If we allow no two particles to coincide
(i.e., impose the hard-core condition), the classical configuration space
for $N$ {\em distinguishable}
particles is ${\cal D}_N= M^N - \Delta$ where 
$\Delta$ is the subset of $M^N$ in which at least two 
points in the Cartesian product coincide.
For indistinguishable particles, we consider
${\cal C}_N = (M^N - \Delta)/S_N $ where $S_N$ is the symmetric
group of $N$ elements. The configuration space, ${\cal C}_N $
or ${\cal D}_N$,
is typically not simply-connected.
Suppose we quantize the theory by using
the path integral formulation. The histories that contribute to the
amplitude for a specified initial configuration to propagate to a
specified final configuration divide up into disjoint sectors
labeled by the elements of the fundamental group
of the configuration space ($\pi_1( {\cal C}_N)$ or
$\pi_1( {\cal D}_N)$) . We have the
freedom to weight the contributions from the different sectors with
different factors, as long as the amplitudes respect the principle
of conservation of probability. It can be shown that this requirement
is equivalent to restricting the weighting factors to be unitary
representations of $\pi_1( {\cal C}_N)$ or $\pi_1( {\cal D}_N)$ .

Let us consider the special cases discussed earlier. A single vortex
moving on a plane has a contractible configuration space and the
quantum mechanics is, therefore, equivalent to that of a free particle.

For two distinguishable vortices on a plane, the fundamental
group of the configuration space, $\pi_1( {\cal D}_2) =Z$,
the set of integers. Notice that
the monodromy eigenstates, $\chi_l$, are irreducible representations
of $Z$ with eigenvalues, $e^{2 \pi i l/n}$. Similar arguments
apply to the cases of two indistinguishable vortices or
two dyons on a plane.\footnote{Incidentally,
with our knowledge of the exact wavefunctions for
two indistinguishable vortices, the second virial coefficient
of non-Abelian vortices can be readily obtained.} The case
of a vortex on a cylinder is topologically equivalent to two vortices
on a plane.

The configuration space of a vortex on a torus is simply the torus 
itself, which has two non-contractible loops. Its fundamental 
group is, therefore, $Z \times Z$. Curiously, we see that only 
the trivial representation is realized in this case. This is true 
on a torus not just for a vortex, but also for a charged particle 
or a flux-charge composite (i.e., a dyon).\footnote{This conclusion 
is based on our assumption that our two-dimensional theory is 
fundamental and the Gauss law constraint is strictly satified. If, 
instead, we regard the two-dimensional theory as an effective theory 
of say a confined electron moving a two-dimensional thin film with 
periodic boundary conditions, there is no reason to impose such a 
strong assumption and the electron wavefunction may transform 
non-trivially under $Z \times Z$. (So long as the field strength 
vanishes, it remains true that the winding around the loop 
$\alpha \beta \alpha^{-1} \beta^{-1}$ leads to no observable 
changes.)}

We construct exact wavefunctions for two vortices on
a plane and a vortex on a cylinder or a torus.
Our success is partly due to the fact that the
fundamental groups are Abelian in these cases.
Any more complicated systems
such as three vortices on a plane or two vortices on a cylinder
or a torus involve configuration spaces with non-Abelian fundamental
groups. In those cases, our simple arguments are doomed to fail.
More powerful methods, yet to be developed, are needed for
tackling those problems. Some of
the results
presented in this manuscript
have also been obtained independently by Brekke, Collins and
Imbo\cite{Brekke}.

We thank J. Preskill for introducing us to the subject
of discrete gauge theories. Helpful discussions with S. Adler, T. D. Imbo,
M. Bucher, H.-F. Chau and P. McGraw are also
gratefully acknowledged. This work was supported in part by
DOE DE-FG02-90ER40542.

\begin{figure}
\caption{An arbitrary base point $x_0$ common to all vortices
and a standard path $\gamma_i$, based at $x_0$, around each vortex.
The case of three vortices is shown in the figure.}
\label{fig1}
\end{figure}
\begin{figure}
\caption{(a) Standard paths $\gamma_1$ and $\gamma_2$, based
at $x_0$, that enclose vortex $1$ and $2$. (b) Vortex $1$ winds around
vortex $2$. (c) Path that, during the winding of vortex $1$ around
vortex $2$, gets dragged to $\gamma_1$. (d) Path that gets dragged
to $\gamma_2$ during the winding.}
\label{fig2}
\end{figure}
\begin{figure}
\caption{(a) Path $\gamma$ that encloses the vortex and path $\alpha$ that
is non-contractible on a cylinder. The vortex winds around the
cylinder. (b) Path $\gamma$ gets dragged to $\gamma'$
during the winding.}
\label{fig3}
\end{figure}
\begin{figure}
\caption{(a) Paths $\alpha$ and $\beta$ are non-contractible loops
on a torus. Path $\gamma$ that encloses the vortex. The vortex winds
around the $\alpha$ loop. (b) Paths $\beta$ and $\gamma$ get dragged to
$\beta'$ and $\gamma'$, during the winding.}
\label{fig4}
\end{figure}
\end{document}